\documentclass[preprint,review,12pt]{elsarticle}

\usepackage{amssymb, subcaption, siunitx, amsmath}

\usepackage{lineno}

\journal{Nuclear Physics A}

\begin{document}


\begin{frontmatter}



\title{Millisecond burst extractions from synchrotrons using RF phase displacement acceleration }


\affiliation[inst1]{organization={CERN},
            city={Geneva},
            postcode={1211}, 
            country={Switzerland}}
            
\affiliation[inst2]{organization={University of Oxford},
            addressline={Keble Road}, 
            city={Oxford},
            postcode={OX1 3RH}, 
            country={UK}}

\author[inst1,inst2]{Pablo A. Arrutia Sota \corref{email}}
\author[inst2]{Philip N. Burrows}
\author[inst1]{Matthew A. Fraser}
\author[inst1]{Francesco M. Velotti}

\cortext[email]{Corresponding author.
E-mail address: pablo.andreas.arrutia.sota@cern.ch}

\begin{abstract}
FLASH radiation therapy calls for the delivery of fast bursted spills of particles with dose delivery times of the order of milliseconds. The requirements overlap with fundamental physics experimental requests that are being studied at CERN, albeit at very different energy scales. In this contribution, a scheme for extracting millisecond bursts from synchrotrons is explored by controlling a third-integer resonant and chromatic extraction with RF phase displacement acceleration. The scheme would be implementable in existing medical and experimental synchrotron facilities. Using a model of the CERN Proton Synchrotron, both single-burst and multi-burst extractions are simulated. Results show that 80 - 90\% of the total beam intensity is extracted in a single burst of 40 - 60 ms. This would correspond to a $\sim$10 ms burst in a typical medical synchrotron, namely the one outlined in the Proton Ion Medical Machine Study. A set of 3 consecutive bursts of 30 ms was simulated in the Proton Synchrotron with optimised machine parameters.

\end{abstract}




\end{frontmatter}

\section{Introduction}

Within the Physics Beyond Colliders (PBC) Study Group at CERN, novel methods of resonant slow extraction from synchrotrons are being explored to satisfy the requirements of future fundamental physics experiments. Recent studies for Enhanced NeUtrino BEams from kaon Tagging (ENUBET) investigated the extraction of millisecond bursts of protons from the CERN Super Proton Synchrotron (SPS) by pulsing the main quadrupole circuit \cite{enubet}. Furthermore, a novel radiation oncology therapy known as FLASH \cite{flash} has awoken interest in the delivery of fast bursted spills of particles in the medical community. FLASH is an ultra-high dose rate irradiation technique that relies on total dose delivery times of the order of milliseconds, often with RF time structure of several MHz. As a consequence, ongoing studies \cite{flash_tech} are evaluating the potential of different accelerator types as candidates for delivery of FLASH therapy. In this paper, an acceleration scheme known as radio frequency (RF) phase displacement \cite{phase} is explored to control the bursting of a resonant third-integer extraction. The same mechanism could also be exploited to feed other resonances like the half-integer, which has been employed in the SPS for fast resonant extractions \cite{half_int} in the past. The flexibility of digital low-level RF control systems in synchrotrons makes this technique easy to implement and optimise. Additionally, no ramping of magnetic elements is required, avoiding magnet ramp rate limits and optics perturbations that could render certain loss reduction techniques such as crystal shadowing \cite{crystal} or octupole folding \cite{octupole} inefficient. RF phase displacement also has the benefit of extracting at constant momentum, while quadrupole sweep techniques produce a time-varying momentum profile. For these reasons, the scheme could be attractive for the exploitation of existing experimental and medical synchrotrons as burst extraction facilities.

RF phase displacement was originally used in the 1960s as a beam acceleration technique in the CERN Intersecting Storage Rings \cite{isr_exp}. In 1999, the Proton-Ion Medical Machine Study (PIMMS) \cite{pimms} considered RF phase displacement as a potential mechanism for slow-extraction spills of around 1 s, but it was ultimately discarded due to the strong modulation it introduced at the repetition period of the sweeping of the RF frequency \cite{ma}. For the burst extraction application, the method's initial weakness is turned into a strength by maximizing the modulation to provide short discrete pulses of particles. Here we study the transverse and longitudinal beam dynamics of the method via simulation; we propose both  single-burst and multi-burst schemes and assess their limits as a burst extraction technique. The CERN Proton Synchrotron is employed as a case study because of its availability for beam tests and its flexible RF equipment that will allow experimental studies in the future. Nevertheless, the procedure can be generalised to any synchrotron and a comparison with SPS and PIMMS parameters is also qualitatively addressed.

\section{RF Burst Extraction Concept}

\subsection{Simulation scheme}

A simple model of the synchrotron was implemented in a code called henontrack \cite{henontrack}, which was exploited to perform tracking simulations. The lattice was reduced to four elements: an effective sextupole, an effective RF cavity, a linear transport matrix and an extraction septum. Particle coordinates were represented in a 4D trace space $(X, X', \phi, \Delta p / p)$ corresponding to normalised horizontal amplitude, normalised horizontal angle, phase offset and relative momentum deviation, respectively. 

\begin{figure}[!ht]
\centering
  \begin{subfigure}[b]{0.75\columnwidth}
    \includegraphics[width=\linewidth]{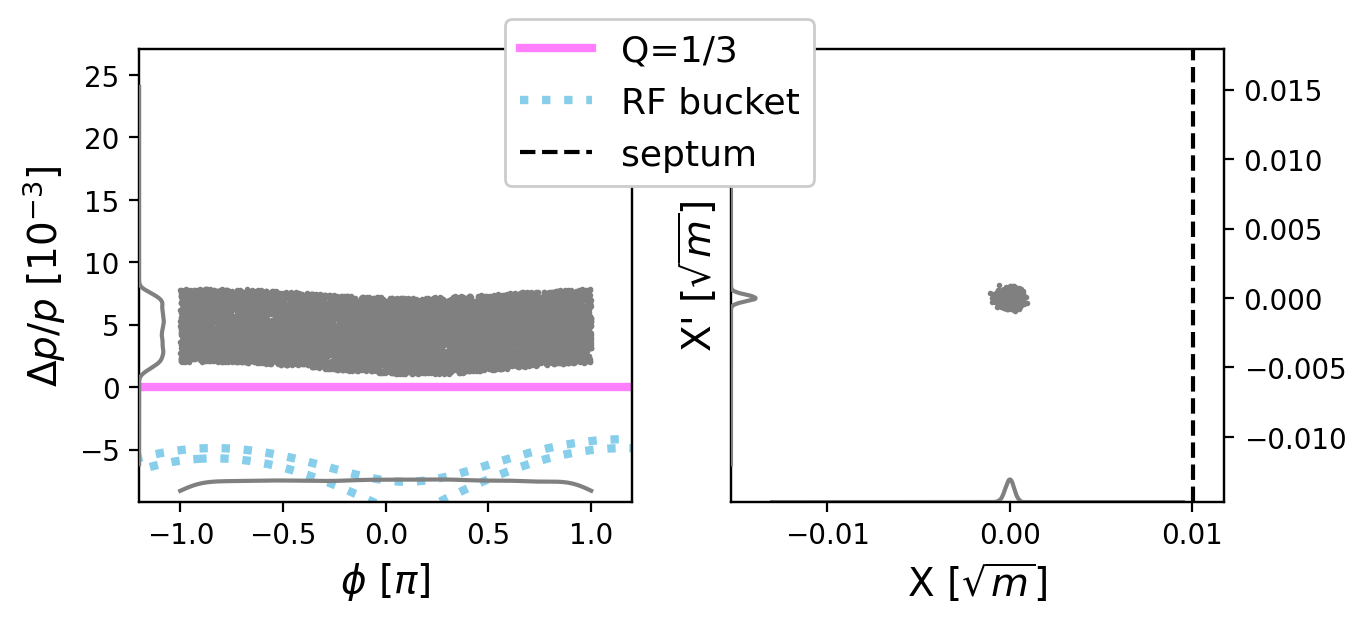}
    \caption{Before frequency sweep.}
    \label{fig: phase_before}
  \end{subfigure}
  \hfill 
  \begin{subfigure}[b]{0.75\columnwidth}
    \includegraphics[width=\linewidth]{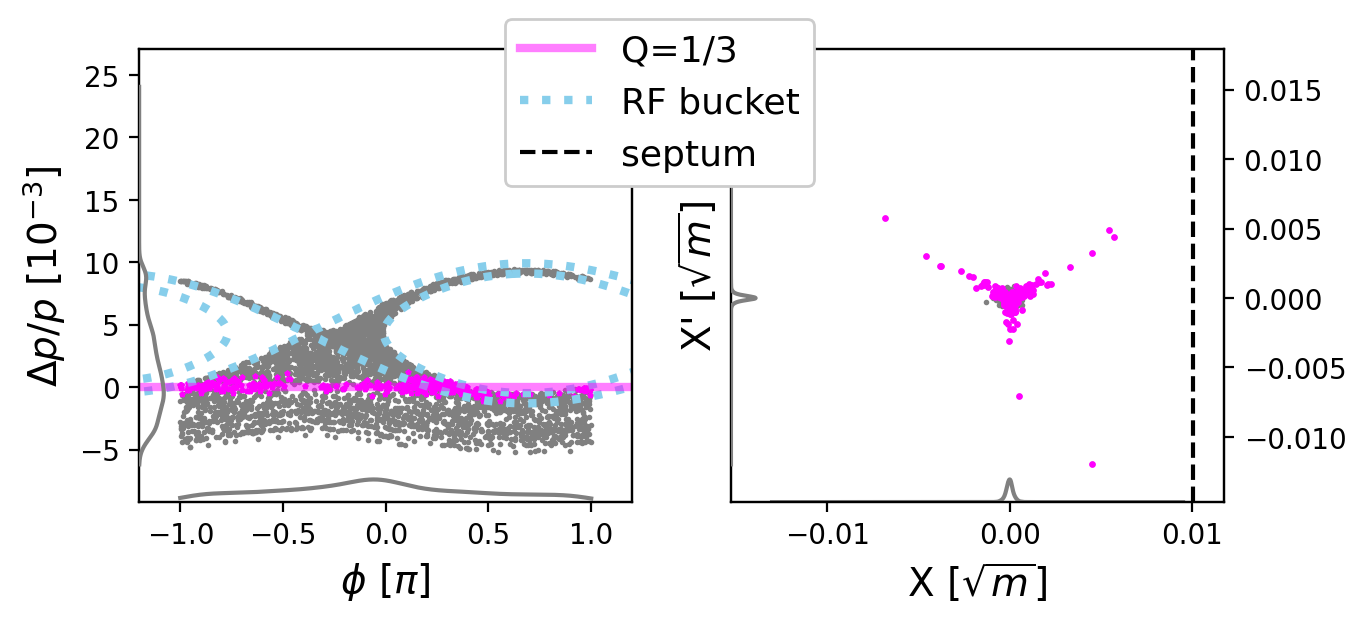}
    \caption{During frequency sweep.}
    \label{fig: phase_during}
  \end{subfigure}
  \begin{subfigure}[b]{0.75\columnwidth}
    \includegraphics[width=\linewidth]{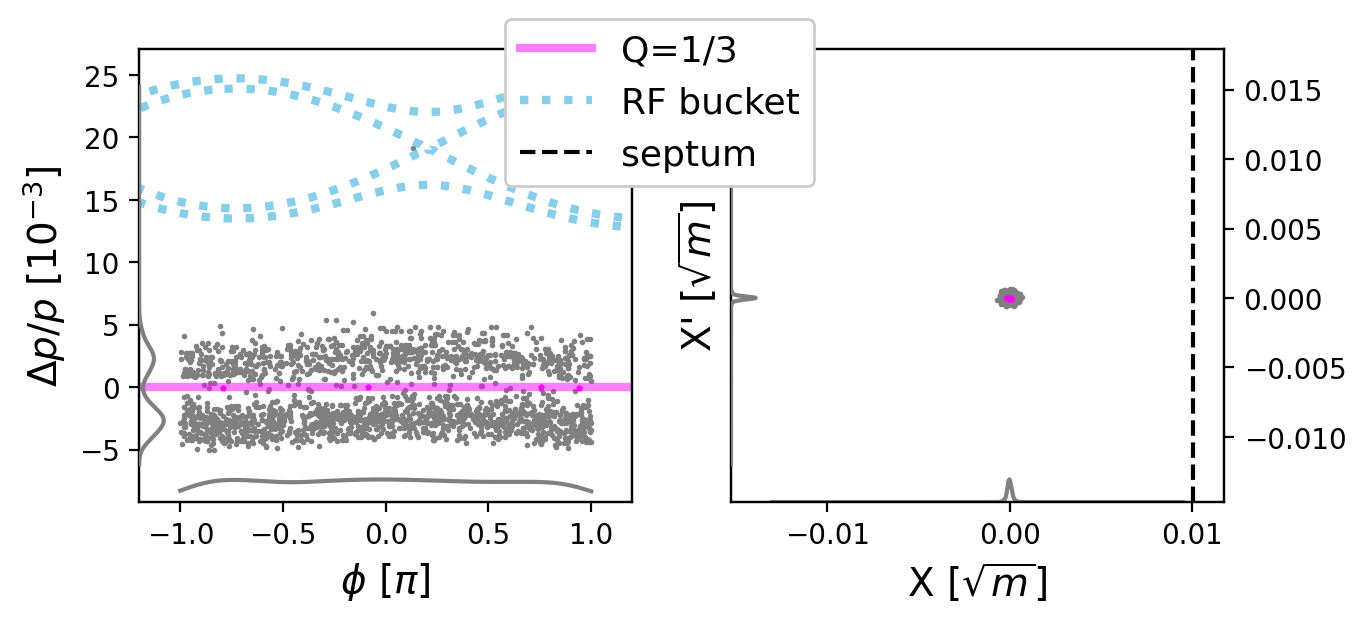}
    \caption{After frequency sweep.}
    \label{fig: phase_after}
  \end{subfigure}
\caption{Longitudinal (left) and transverse (right) phase spaces during a single frequency sweep of RF phase displacement extraction.}
\label{fig: phase_space}
\end{figure}

The RF phase displacement scheme is shown in Fig.~\ref{fig: phase_space} and can be summarised as follows:

\begin{enumerate}
    \item An ensemble of $10^4$ particles is generated with a horizontal transverse Gaussian distribution of geometric RMS emittance $\epsilon_{G, RMS}$ and a uniform longitudinal distribution of momentum spread $(\Delta p / p)_0$ (Fig.~\ref{fig: phase_before}). 
    \item The RF system is programmed to follow a linear frequency ramp that sweeps empty buckets through the stack of particles in longitudinal phase space (Fig.~\ref{fig: phase_during}).
    \item As particles are kicked by the RF cavity they are accelerated, which results in a change in their tune via chromaticity.
    \item Particles with resonant tune gain amplitude until they jump over the septum and are extracted.
    \item After the sweep, some particles may remain in the machine (Fig.~\ref{fig: phase_after}).
    \item Steps 2-4 are repeated $n$ times to provide $n$ bursts.
\end{enumerate}

\begin{figure}[!ht]
\centering
  \begin{subfigure}[b]{0.75\columnwidth}
    \includegraphics[width=\linewidth]{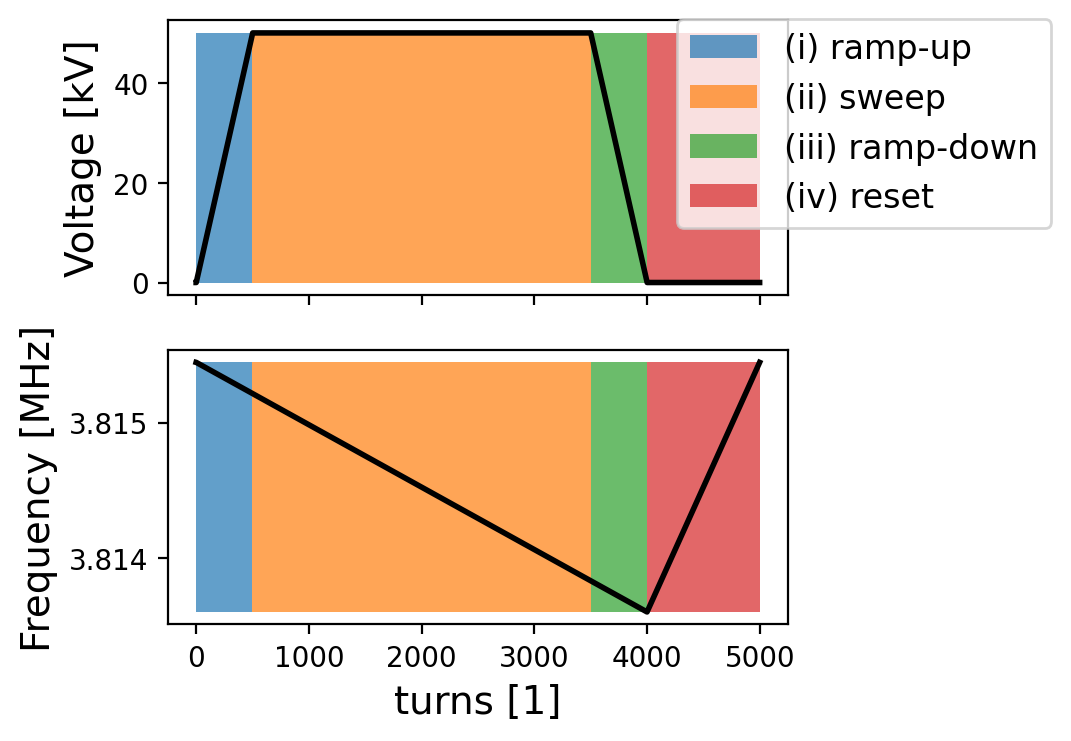}
  \end{subfigure}
\caption{Example of voltage and frequency programs for a single burst: (i) the voltage ramps up to its nominal value while the frequency starts a linear sweep, (ii) the voltage stays flat while the frequency continues its sweep, (iii) the voltage ramps down to zero while the frequency ends its linear sweep and (iv) the voltage stays at zero while the frequency resets to its initial value.}
\label{fig: program_single_burst}
\end{figure}

An example of the voltage and frequency programs employed in simulation is shown in Fig.~\ref{fig: program_single_burst} . For all simulations, the voltage was ramped (up or down) in 500 turns and the frequency was reset to its initial value in 1000 turns. To produce $n$ bursts, $n$ of these programs were concatenated.

\subsection{Relevant parameters}

When the RF frequency is swept through the circulating beam, the distribution of $\Delta p / p$ experiences both an average coherent displacement $\mu$ and a blow-up with root mean square increase in spread $\sigma$. The effect on the $\Delta p / p$ distribution can be derived analytically \cite{scattering,lee} and expressed as,

$$\mu = \frac{A_0}{2\pi} \alpha(\Gamma), \ \ \sigma = \frac{A_0}{2\pi} \Gamma,$$
where $\Gamma=\sin\phi_s$, $\phi_s$ is the synchronous phase, $\alpha \approx \frac{1-\Gamma}{1+\Gamma}$ is the bucket area factor and $A_0$ is the stationary bucket area in ($\phi, \Delta p / p$) space given by,

$$A_0 = 16 \sqrt{\frac{eV}{2\pi\beta^2 E h |\eta|}},$$
where $e$ is the electron charge, $V$ is the RF voltage, $\beta$ is the relativistic speed factor, $E$ is the beam energy, $h$ is the RF harmonic number and $\eta$ is the slip factor. The expressions can be normalised to the initial momentum spread of the beam $(\Delta p / p)_{0}$ to obtain,

\begin{equation}
\bar{\mu} = \frac{\mu}{(\Delta p / p)_{0}} = r \alpha(\Gamma), \ \ \bar{\sigma} = \frac{\sigma}{(\Delta p / p)_{0}} = r \Gamma,
\label{eq: mubar_sigmabar}
\end{equation}
where $r=\frac{A_0}{2\pi(\Delta p / p)_{0}}$ is the ratio between the stationary bucket area and the phase space area of the coasting beam. This relationship provides an intuitive interpretation: to push a fraction $1/n$ of the beam into the resonance, the sweeping bucket must have approximately a factor $1/n$ of the stack area. However, one must additionally account for the blow-up contribution, which makes the relationship less straightforward. Furthermore, $A_0$ and $\Gamma$ vary slightly (1\% for the PS) during the sweep as $E$ changes but $V$ stays fixed, which makes $r$ correct only for the average parameters. Both of these effects were included in the numerical tracking. On the other hand, the variation of $\eta$ as a function of $E$ was not included (since linear longitudinal transport was assumed), but the change is small ($\sim$ 1\% for the PS) if the machine is operated far from transition energy. For a large number of sweeps, the variation in $\eta$ could produce a noticeable perturbation \cite{pimms}.

In practice $\Gamma$ will have a maximum value for a given $V$, since the RF frequency cannot be ramped arbitrarily fast. The ramp rate $\frac{df_{RF}}{dt}$ is given by,

$$\frac{df_{RF}}{dt} = \frac{\eta f_{RF}^2 e V}{\beta^2 E} \Gamma = F(V) \Gamma$$,
where $f_{RF}$ is the RF frequency and $F(V)$ is the proportionality constant between $\Gamma$ and ramp rate in physical units. This establishes a maximum $\Gamma = \Gamma_{max}$ given by,

$$\Gamma_{max} = \frac{1}{F(V)}\frac{df_{RF}}{dt}\bigg|_{max}$$,
where $\frac{df_{RF}}{dt}|_{max}$ is the maximum RF ramp rate. This directly sets a minimum $r=r_{min}=r(\Gamma_{max})$.

\subsection{Longitudinal and transverse transit times}

In the longitudinal plane, a timescale for each burst can be estimated by ignoring dynamic effects; simply the time needed to move the entire bucket area across the same amount of area in the stack. In other words, the (virtual) synchronous particle inside the sweeping bucket must travel a distance $\Delta p / p = \bar{\mu}(\Delta p / p)_0 + 2H$, where $H$ is the bucket height. Ultimately, numerical simulations are needed to account for dynamic effects. The timescale $\tau_L$ (in number of turns) is given by,

\begin{equation}
\tau_L = \frac{[1 + \frac{\pi}{2} r Y(\Gamma)]\bar{\mu}(\Gamma)}{\Gamma}
\bigg[ \frac{(\Delta p / p)_0}{\frac{eV}{\beta^2 E}} \bigg],
\label{eq: tau_l}
\end{equation}
where $Y(\Gamma)$ is the bucket height factor \cite{lee}. The first term captures the dependence on $\Gamma$ and the second term provides the time needed for the synchronous particle to cross the entire stack at a given voltage. The latter can easily be compared across machines by setting $V$ to the maximum voltage $V_{max}$, as shown in Table \ref{tab: machine_params} by $\tau_{L, 0}$. 

In the transverse plane, particles take a finite amount of time to gain amplitude and jump over the extraction septum. This transit time varies between particles and it is dependent on their initial coordinates when entering the resonant region and their tune speed \cite{pimms}. Nevertheless, a characteristic timescale $\tau_T$ (in number of turns) can be obtained \cite{cornelis} and expressed as,

$$\tau_T = \frac{8}{S \sqrt{\epsilon_{G, RMS}}},$$
where $S$ is the virtual sextupole strength of the machine \cite{pimms}. Dynamic effects faster than $\tau_T$ will limit the extraction process, e.g. certain particles could cross the resonant region too fast to be extracted. In RF phase displacement the coherent kicks from the RF could make the tune speed of certain particles too large, transporting them back to a stable tune before they can reach the septum. Numerical simulations quantify the severity of this effect.

\begin{table}[hbt]
\centering
\begin{tabular}{||l c c c||} 
 \hline
 Parameter & PS,  & PIMMS & SPS \\ [0.5ex] 
   & h=8 (16) &  &  \\ [0.5ex] 
 \hline\hline
 Kinetic E. [GeV] & 24 & 0.25 & 400\\
 \hline
 $\epsilon_{G, RMS}$ [mm.mrad] & 0.059 & 0.67 & 0.0018 \\ 
 \hline
 T [$\mu$s] & 2.1 & 0.4 & 23\\  
 \hline
 h [$1$]& 8 (16) & 1 & 4620\\
 \hline
 $V_{max}$ [kV] & 200 & 4 & 10000\\ 
 \hline
 $\eta$ [$1$] & 0.024 & -0.37 & 0.0018\\
 \hline
 S [m$^{-1/2}$] & 77 & 30 & 170\\ 
 \hline
 $F(V_{max})$ [kHz/ms] & 0.37 (0.74)  & 18 & 0.40 \\
 \hline
 $(\Delta p / p)_0$ [$10^{-3}$] & 6 & 4 & 3\\
 \hline
 $\tau_T$ [turns] & 430 & 330 & 350\\
 \hline
 $r_{max}$ [$1$] & 1.1 (0.78) & 1.3 & 0.58\\
  \hline
 $\tau_{L, 0}$ [turns] & 720 & 450 & 120\\[1ex] 
 \hline
\end{tabular}
\caption{Relevant machine parameters for RF phase displacement burst extraction.}
\label{tab: machine_params}
\end{table}

Table \ref{tab: machine_params} summarises the relevant parameters for the PS (both for $h=8$ and $h=16$), the SPS and a PIMMS-like synchrotron. The PIMMS-like machine and the PS ($h=8$) differ considerably in nominal parameters, but the normalised parameters $\tau_T$, $\tau_{L, 0}$ and $r$ are within 40\% of each other. Therefore, we expect results in the PS to be indicative of the PIMMS-like machines, as long as they are expressed in terms of unitless parameters. Due in part to its high RF harmonic number, the SPS parameters differ from the PS parameters.

\section{Simulation Results and Discussion}

\subsection{Single burst extraction}

\begin{figure}[t!h]
\centering
  \begin{subfigure}[b]{0.75\columnwidth}
    \includegraphics[width=\linewidth]{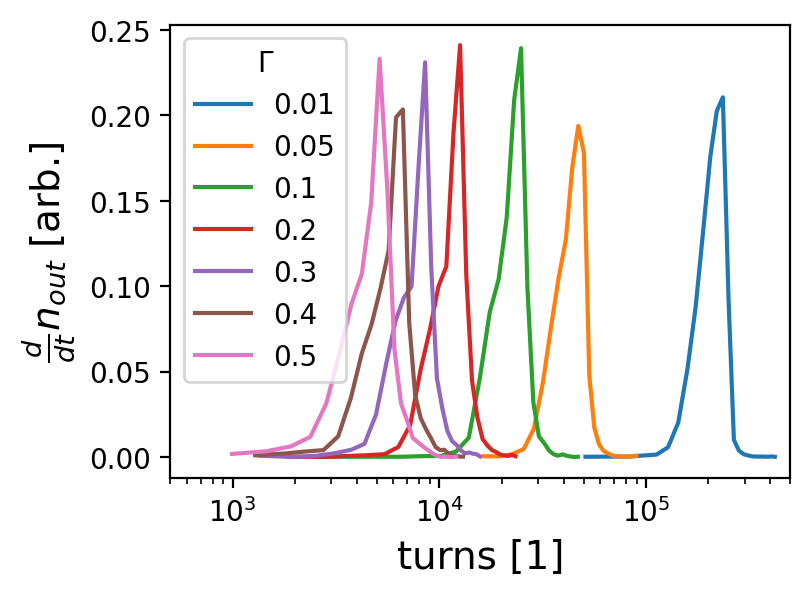}
    \caption{Instantaneous extracted particle number.}
  \end{subfigure}
  \begin{subfigure}[b]{0.75\columnwidth}
    \includegraphics[width=\linewidth]{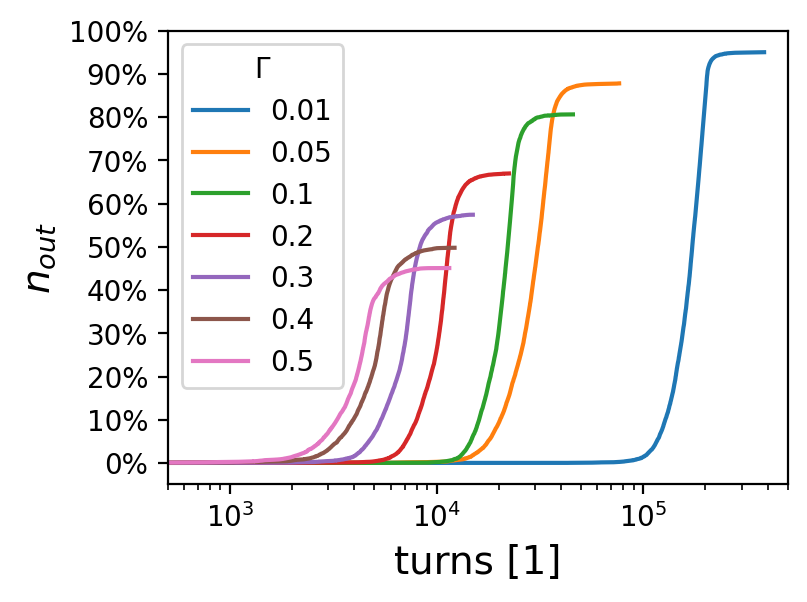}
    \caption{Cumulative extracted particle number.}
  \end{subfigure}
\caption{Extracted particles $n_{out}$ for single burst scheme with different $\Gamma$.}
\label{fig: spills}
\end{figure}

Single burst simulations were performed for a fixed voltage $V=V_{max}$, and fixed harmonic $h=8$ (as it provided a larger $r$), whilst scanning $\Gamma$. The instantaneous and cumulative intensities of the extracted spills are shown in Fig.~\ref{fig: spills}.  As predicted by Eq.~\ref{eq: tau_l}, the larger the $\Gamma$, the shorter the bursts (shorter $\tau_L$), but fewer particles are extracted as predicted by Eq.~\ref{eq: mubar_sigmabar} because of the smaller $\bar{\mu}$.

\begin{figure}[tb]
\centering
  \begin{subfigure}[b]{0.6\columnwidth}
    \includegraphics[width=\linewidth]{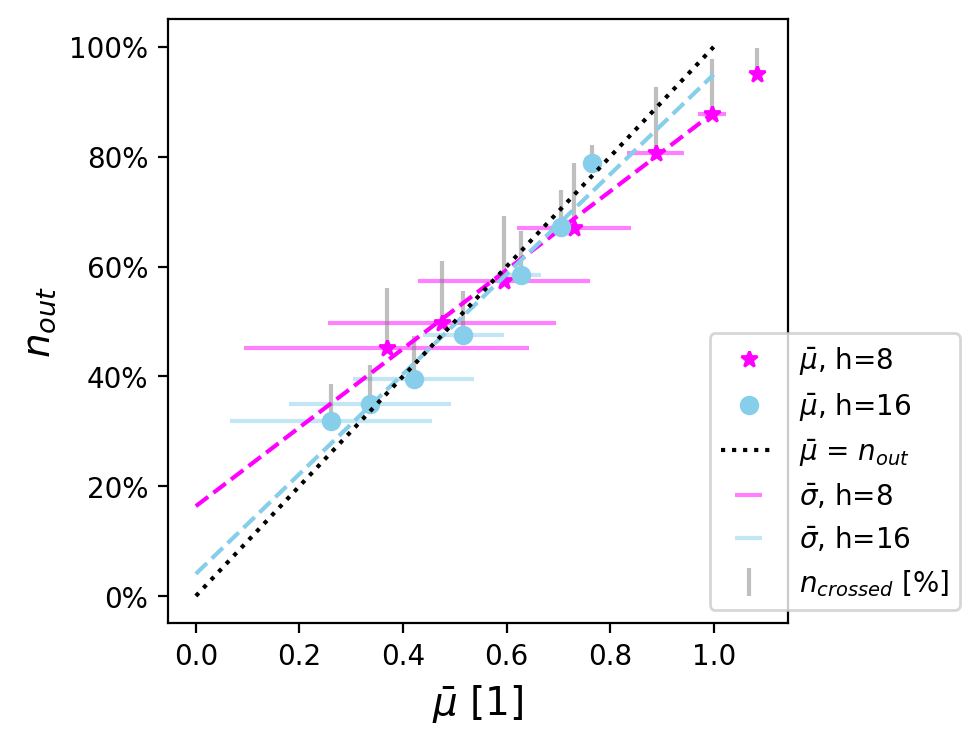}
  \end{subfigure}
\caption{Extracted particles $n_{out}$ v. $\bar{\mu}$. The dashed lines show linear fits to the simulation data. The horizontal lines ($\bar{\sigma}$) indicate the size of the momentum blow-up effect calculated from Eq.~\ref{eq: mubar_sigmabar}. The grey vertical lines show the portion of particles $n_{crossed}$ that crossed the resonance but did not get extracted. They are only added in the higher $n_{out}$ direction since they indicate the additional percentage of particles that would have been extracted if the extraction happened instantaneously when entering the resonance (i.e. they express the asymptotic limit when $\tau_T = 0$).}
\label{fig: nout}
\end{figure}

Simulations were repeated with identical $V$ and $\Gamma$, but with $h=16$. This changes $r$ and $\tau_L$ while keeping all other parameters constant. Fig.~\ref{fig: nout} shows the percentage of extracted particles $n_{out}$ as a function of $\bar{\mu}$. In an idealised situation without blow-up or finite extraction time, the simulated data points would lie on the $\bar{\mu}=n_{out}$ line. At low $\bar{\mu}$ blow-up effects drive a substantial number of particles into or away from the resonance due to the large $\bar{\sigma}$. For all settings, a certain number of particles experience large changes in tune per turn. This fast change in tune has the potential to increase the transverse emittance of the extracted beam \cite{javier} and some particles $n_{crossed}$ even cross the resonance region too quickly to be extracted, becoming non-resonant again and remaining in the machine. This reduces the efficiency of the scheme but could be combated by shortening $\tau_T$, as particles would then need less turns to reach the septum. These effects are included in all simulation results. Nevertheless, the simulated dependence of $\bar{\mu}(n_{out})$ remains approximately a linear trend (dashed lines), and especially for $h=16$, where the blow-up effects are smaller due to the smaller $r$. 

\begin{figure}[!th]
\centering
  \begin{subfigure}[b]{0.6\columnwidth}
    \includegraphics[width=\linewidth]{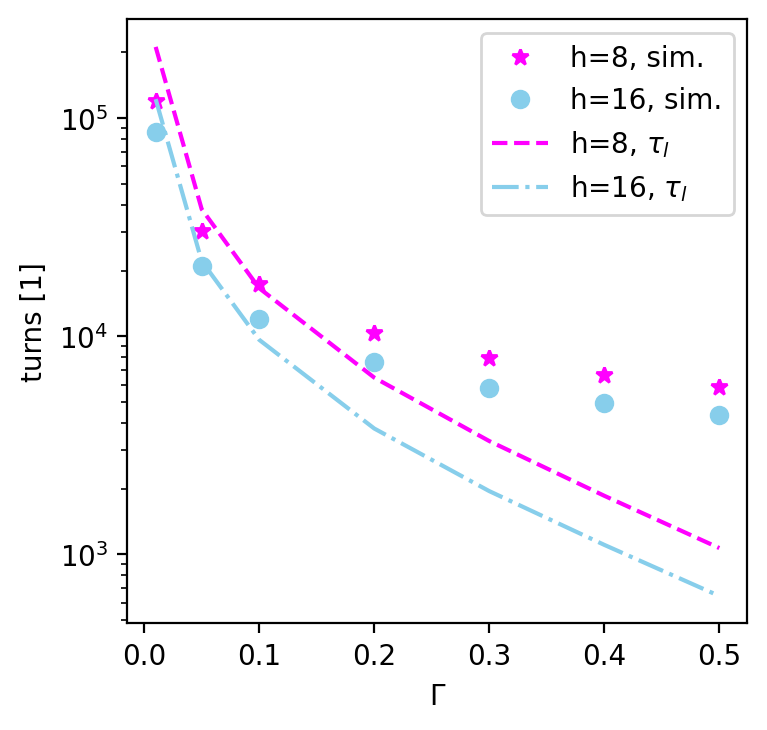}
  \end{subfigure}
\caption{Extraction time v. $\Gamma$. The dashed lines show $\tau_L$ as a function of $\Gamma$.}
\label{fig: tau_l}
\end{figure}

Fig.~\ref{fig: tau_l} shows the extraction time as a function of $\Gamma$. The extraction time was computed as the number of turns required to extract from 1\% to 99\% of the total extracted intensity, i.e. including only the central 98\%. This metric was chosen to stop outliers from dominating the calculation, as they can dramatically increase the total extraction time (by a factor 2 or more). $\tau_L$ is within a factor 2 of the simulation output for all the data points. However, dynamic effects clearly play an important role at large $\Gamma$.

\subsection{Multiple burst extraction}

\begin{figure}[!ht]
\centering
  \begin{subfigure}[b]{0.75\columnwidth}
    \includegraphics[width=\linewidth]{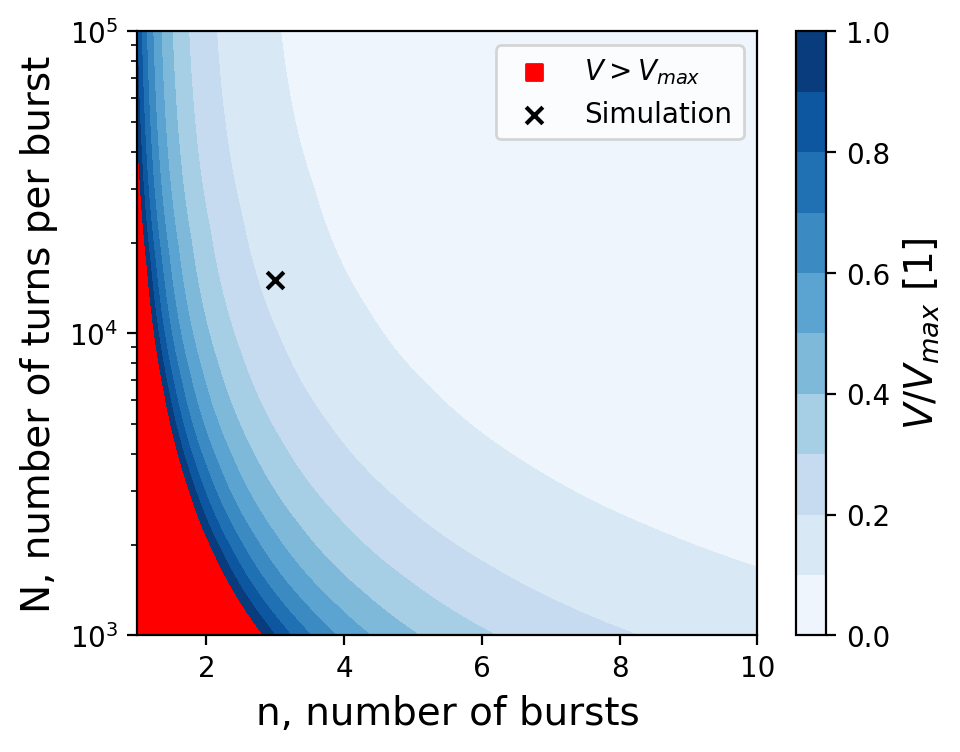}
  \end{subfigure}
\caption{Dependence of $V$ on $n$ and $N$ for the PS, $h=8$.}
\label{fig: V_vs_nN}
\end{figure}

\begin{figure}[!ht]
\centering
  \begin{subfigure}[b]{0.75\columnwidth}
    \includegraphics[width=\linewidth]{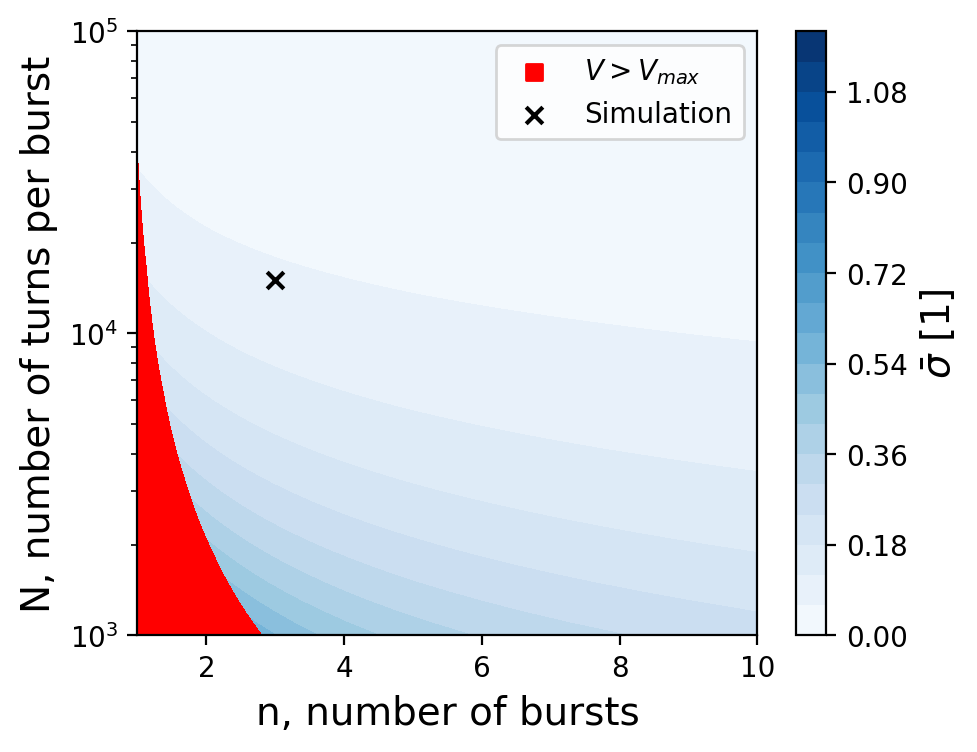}
  \end{subfigure}
\caption{Dependence of $\bar{\sigma}$ on $n$ and $N$ for the PS, $h=8$.}
\label{fig: sigmabar_vs_nN}
\end{figure}

One can exploit the knowledge acquired in the previous subsection to generalise the scheme to multiple bursts, which is highly relevant for FLASH therapy. For a given set of machine parameters, one can estimate an initial guess for $(V, \Gamma)$ such that the extraction has $n$ bursts with $N$ turns per burst by solving the system of equations,

\begin{equation}
    \begin{cases}
    \ \bar{\mu}(V, \Gamma) = \frac{1}{n}\\ 
    \ \tau_L(V, \Gamma) = N, 
    \end{cases}
\end{equation}

as shown in Figs.~\ref{fig: V_vs_nN} and  \ref{fig: sigmabar_vs_nN}. Such an estimate does not account for blow-up and finite transit time and would have to be further studied and optimised. The scheme would be limited by $V_{max}$ and $\tau_T$. The latter could be reduced with higher $S$ or larger $\epsilon_{G, RMS}$. 

Fig.~\ref{fig: V_vs_nN} shows the required voltage for different combinations of $n$ and $N$, which establishes the forbidden region due to voltage limitations. Furthermore, larger $\Gamma$ leads to larger $\bar{\sigma}$, which may complicate operation and optimisation of consecutive bursts. Figure \ref{fig: sigmabar_vs_nN} shows $\bar{\sigma}$ for different combinations of $n$ and $N$.

\begin{figure}[!ht]
\centering
  \begin{subfigure}[b]{0.65\columnwidth}
    \includegraphics[width=\linewidth]{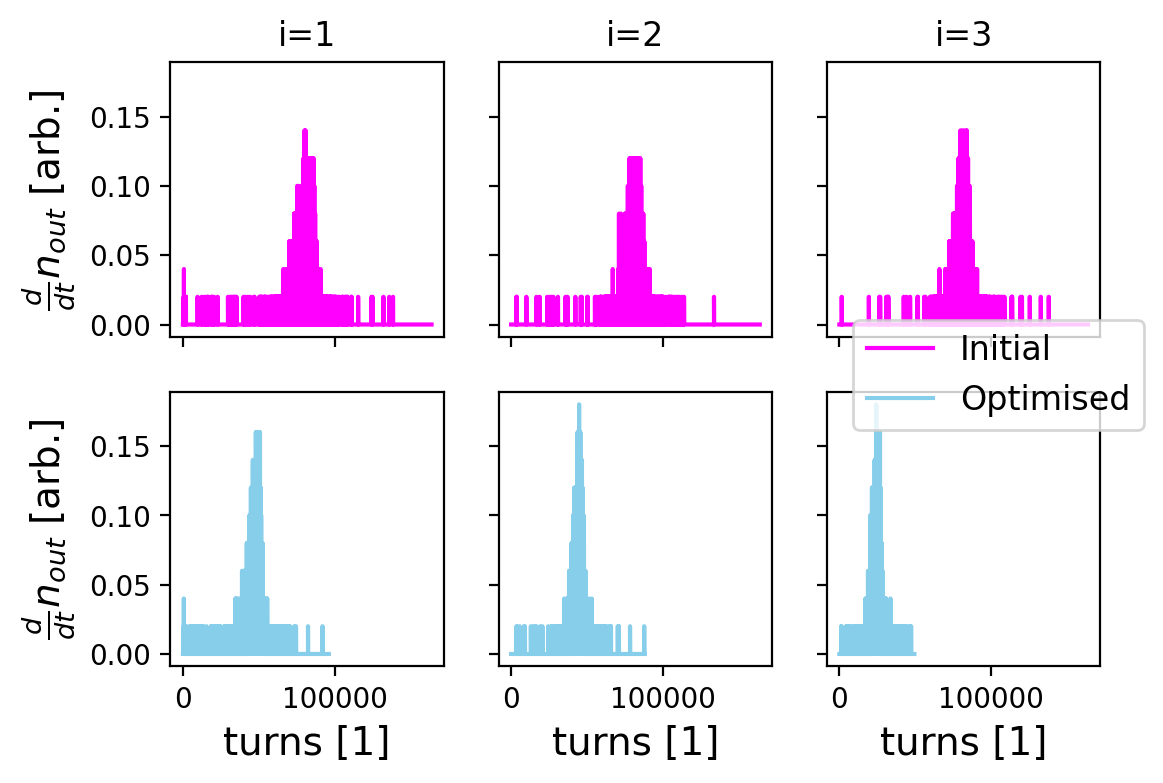}
  \end{subfigure}
\caption{Instantaneous intensity profiles for multiple burst scheme in the PS ($h=8$) with $n=3$ and $N=15000$.}
\label{fig: bursts_initial_optimised}
\end{figure}

\begin{figure}[!ht]
\centering
  \begin{subfigure}[b]{0.65\columnwidth}
    \includegraphics[width=\linewidth]{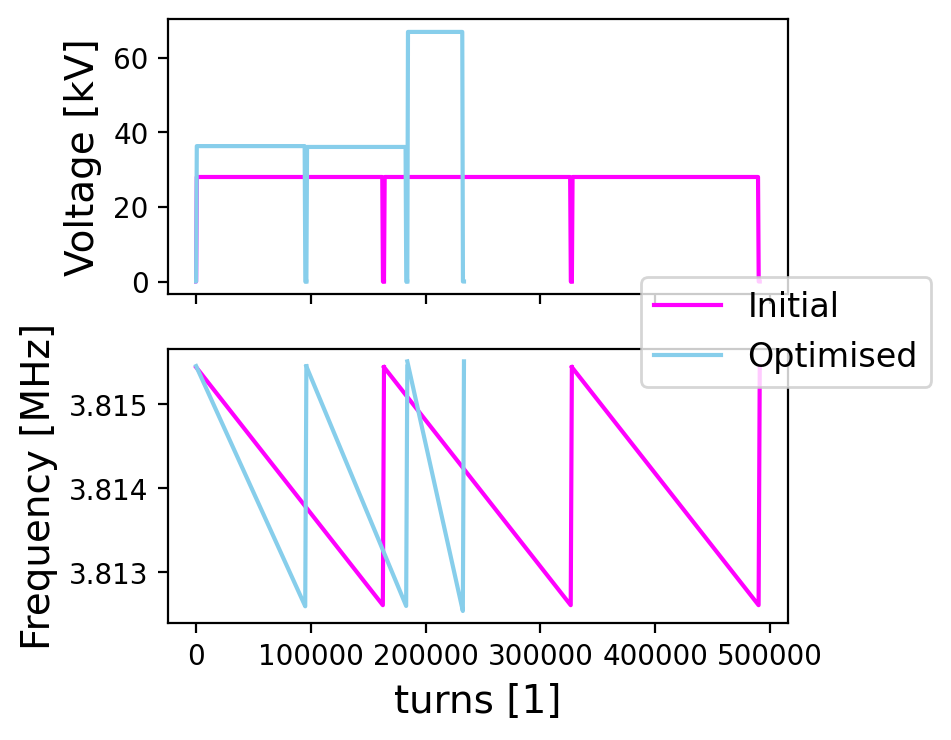}
  \end{subfigure}
\caption{Voltage (top) and frequency (bottom) programs for multiple burst scheme in the PS ($h=8$) with $n=3$ and $N=15000$.}
\label{fig: program_multi_burst}
\end{figure}

To test the validity of the approximations made in Figs.~\ref{fig: V_vs_nN} and \ref{fig: sigmabar_vs_nN} tracking simulations were carried out for the specific combination of $n=3$ and $N=15000$ (chosen as an example). The settings from the contour plots were used as initial conditions for an optimisation procedure employing the algorithm Py-BOBYQA \cite{boby}. In order to account for $n_{crossed}$, the target $n_{out}$ was set to 90\% of the total intensity. This lead to an initial guess of $V=$28 kV, $\Gamma=0.16$ and an expected blow-up of $\bar{\sigma}=0.06$. The optimiser was allowed to vary $V$ and $\Gamma$ for each sweep independently. The cost function was specified as,

$$C = \sqrt{\frac{\sum_i (n_{out, i} - 1/3)^2}{(1/3)^2} + \frac{\sum_i (N_i - 15000)^2}{15000^2}},$$
where $i=1,2,3$ is the sweep number and $n_{out, i}$ was normalised to 90\% of the total intensity. Fig.~\ref{fig: bursts_initial_optimised} shows the instantaneous intensity profiles for both the initial guess and the optimised settings found by the optimiser: $V=(36, 36, 67)$ kV, $\Gamma = (0.21, 0.23, 0.24)$. Fig.~\ref{fig: program_multi_burst} shows the voltage and frequency programs for both schemes.    

\begin{figure}[!ht]
\centering
  \begin{subfigure}[b]{0.65\columnwidth}
    \includegraphics[width=\linewidth]{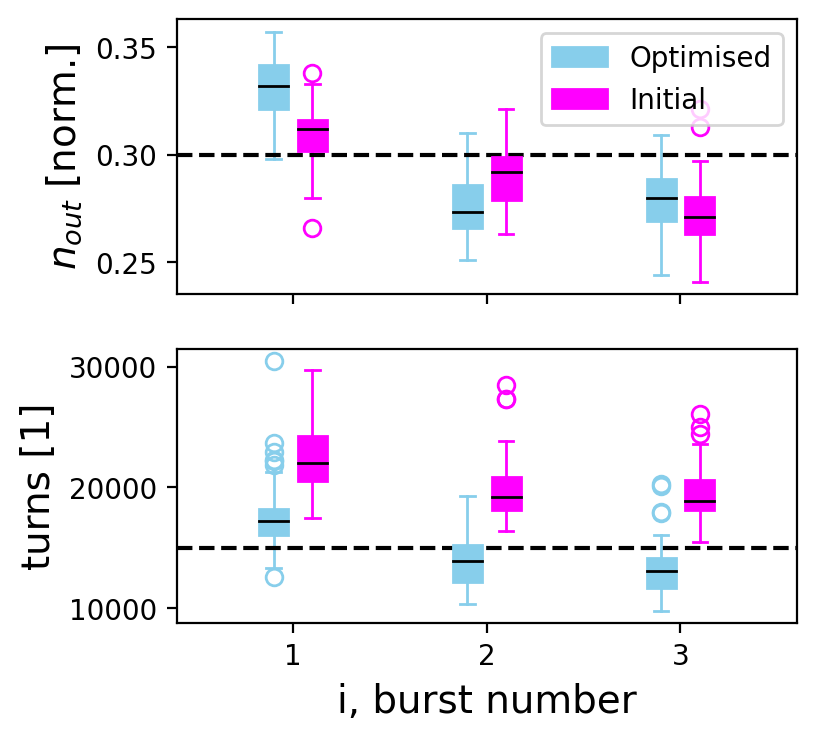}
  \end{subfigure}
\caption{Extracted intensity per burst (top) and burst turns (bottom) for multiple burst scheme in the PS ($h=8$) with $n=3$ and $N=15000$. The dashed lines show target values.}
\label{fig: multi_burst}
\end{figure}

The simulation was executed fifty times with $10^3$ particles to test the robustness of the scheme and the results are shown in Fig.~\ref{fig: multi_burst}. The initial setting produced longer extraction times than the target for all bursts, which is consistent with Fig.~\ref{fig: tau_l}; $\tau_L$ consistently underestimates the extraction time for $\Gamma > 0.1$. The optimised settings brought the extraction time down reducing the average relative error in N from 37\% to 12\%. The average relative error in $n_{out}$ increased from 5\% to 9\%, but since the cost function weighs both errors equally this solution is preferred. A different cost function could be more suitable depending on the specific operational requirements. Furthermore, the solution to this cost function is likely to be non-unique and further optimisation could be possible (e.g. by including more sophisticated control on the voltage and frequency programs). This particular implementation is a simple showcase of the flexibility of the scheme.

\section{Conclusion and Outlook}

RF phase displacement acceleration was exploited to provide short bursts of particles from a synchrotron in a scheme that could be attractive for exploitation at existing experimental and medical synchrotrons, with FLASH therapy in mind. Employing a simplified model of the PS, it was shown that 80 - 90\% of the total beam intensity could be extracted in a single burst of 20000 - 30000 turns. This corresponds to 40 - 60 ms in the PS and 8 - 12 ms in PIMMS-like machines. Moreover, a few basic parameters were computed to characterise the extraction scheme. The parameters were informative, but simulation showed that dynamic effects ultimately played an important role in the extracted intensity and extraction time. A 3-burst scheme with 15000 turns per burst could be realised in a simulation of the PS with optimised machine parameters. Future work will demonstrate the implementation of the extraction technique with measurements at the PS, whilst exploiting sophisticated optimisation and control tools.

\section{Acknowledgements}

This work was supported by the Physics Beyond Colliders (PBC) Study Group. The authors thank PBC for their support. The same gratitude is extended to H. Damerau and M. Vadai from the CERN SY-RF team for the enriching discussions on RF systems. The first author would also like to thank the CERN Accelerator Beam Transfer group for supporting the attendance to the FLASH Radiotherapy \& Particle Therapy conference in 2021.


 \bibliographystyle{elsarticle-num} 
 \bibliography{main}





\end{document}